\documentclass[11pt,a4paper]{article}

\usepackage{amsmath}
\usepackage{amssymb}
\usepackage{amsthm}
\usepackage[mathscr]{euscript}
\usepackage{graphicx}
\usepackage{wrapfig}
\usepackage{caption}
\usepackage{subcaption}
\usepackage{booktabs}
\usepackage[bookmarks=true,bookmarksopen=true,colorlinks=true,breaklinks=true,linkcolor=blue,citecolor=blue]{hyperref}

\newcommand{\Rc}{R_{\mathrm{circ}}}

\newcommand{\dz}{\,\mathrm{d}z}

\newcommand {\md} {\mathrm d}

\newcommand {\J} {\mathcal{J}}
\newcommand {\M} {\mathcal{M}}

\newcommand{\Eqref}[1]{Eq.~\eqref{#1}}

\newcommand{\Sectionref}[1]{Section~\ref{#1}}

\newcommand{\Figref}[1]{Figure~\ref{#1}}

\begin{document}

\title{Stationary solutions of the axially symmetric Einstein-Vlasov system: present status and open problems}
\author{
  Ellery Ames\\
  Department of Mathematics \\
  California Polytechnic University, Humboldt\\
  Arcata, California 95521, USA \\
  email: ellery.ames@humboldt.edu
  \\ \\
  H{\aa}kan Andr{\'e}asson\\
  Mathematical Sciences\\
  Chalmers University of Technology\\
  University of Gothenburg\\
  SE-41296 Gothenburg, Sweden\\
  email: hand@chalmers.se
}
        
\date{\today}
\maketitle

\begin{abstract}
The purpose of this work is to review the status about stationary solutions of the axially symmetric Einstein-Vlasov system with a focus on open problems of both analytical and numerical nature.
For the latter we emphasize that the code used to construct stationary solutions in \cite{Ames2016,Ames2019} is open source, see \cite{Ames2023joss}.
In the analytical setting the open problems include establishing methods for proving existence of axisymmetric stationary solutions which are far from spherically symmetric, both in the general case and for certain special classes of solutions pointed out in the text. 
In the numerical setting there are intriguing properties of highly relativistic solutions that demand further attention, such as whether a sequence of such stationary solutions can approach a Kerr black hole, or if they necessarily approach the thin ring limit reminiscent of cosmic strings. 
The question of whether stationary solutions include states with thin-disk like morphologies as seen in many galaxies is also open. 
Finally, there are opportunities to extend this research to new settings such as the case of massless particles and coupled black hole-matter systems.
We believe that some of the open problems highlighted here are of central importance for the understanding of nature.
\end{abstract}

\section{Introduction}
\label{sec.introduction}
There are many reasons why studying the axially symmetric Einstein equations coupled to matter is of great interest. In 1955 Wheeler wrote in his seminal work on geons \cite{Wheeler1955}: 
\begin{quote}
The simple toroidal geon forms the most elementary object of geon theory much as a simple circular orbit constitutes the first concept of planetary theory. But the simplest physics does not go in the geon case with the simplest mathematics. 
\end{quote}
Wheeler here expresses the wish to study his ideas on geons in the axially symmetric case but due to the complicated nature of the equations in that case he instead invented a spherically symmetric toy model which he called ``Idealized spherically symmetric geons". Although the research field has made tremendous progress since this work was published, analytic results about Einstein's equations coupled to matter for isolated bodies are still very limited in the axially symmetric case. 
The development of numerical solutions has perhaps been even more dramatic since Wheeler's work from 1955  (where Wheeler in fact solved the Einstein equations numerically in the spherically symmetric case) but  numerical studies about Einstein's equations coupled to matter often still concern toy models for the matter such as dust or a scalar field.
There are of course exceptions, but at least for the Einstein-Vlasov system only a few investigations have been carried out.
Shapiro and Teukolsky were pioneers and initiated an important study in the early 90's on the axisymmetric Einstein-Vlasov system.
They considered both the dynamical case \cite{Shapiro1991,Shapiro1992a,Shapiro1992b} and the stationary case \cite{Shapiro1993a,Shapiro1993b}.
More recently this line of research has been continued in \cite{Ames2016,Ames2019,Ames2021,Ames2023}. 
Many  questions have been answered in these works but several central questions remain open. 

The purpose of the present contribution is to review the status about stationary solutions of the axially symmetric Einstein-Vlasov system, including both analytical and numerical work, and in particular to discuss open problems.
We hope this discussion will stimulate activity in the field so that progress can be made on central problems.
The open source code GECo \cite{Ames2023joss}, or further development of this code, may be useful in addressing some of the open problems discussed below.

In order to put the discussion below into context, in the remainder of this section we briefly discuss the spherically symmetric setting from a numerical perspective in \Sectionref{sec.introduction.spherical_symmetry}, and in \Sectionref{sec.introduction.axisym_ev_system} include some equations and concepts from the axially symmetric literature.
For a general background on the Einstein-Vlasov system we refer to \cite{Andreasson2011,Rendall1996,Sarb2022}.
In \Sectionref{sec.survey} we briefly summarize the known results in the field. 
Further discussion of results in the context of open problems is presented in \Sectionref{sec.discussion}. 
Finally in \Sectionref{sec.evolution} we make a few remarks on the evolutionary setting.

\subsection{Comparison of numerical methods in the spherically symmetric setting}
\label{sec.introduction.spherical_symmetry}
Spherically symmetric static solutions to the Einstein-Vlasov system are well-studied both analytically and numerically, cf. \cite{Andreasson2011}.
There is a crucial difference in how solutions in the spherically symmetric case and solutions in the axially symmetric case are obtained.
In the former case the equations can be solved as an initial value problem in the sense that given data at $r=0$, a non-linear system of integro-differential equations can be solved straightforwardly, at least numerically.
In this way a large variety of solutions can be constructed.
An example is the class of highly relativistic multi-peak solutions that was discovered in \cite{Andreasson:2007ix}.
These solutions are believed to be unstable, but still they are easy to obtain numerically. 

In the axially symmetric case the situation is different.
The numerical methods that have been used in previous works \cite{Shapiro1993a,Shapiro1993b,Ames2016,Ames2019} rely on an iteration procedure which must converge in order to produce a solution.
Such iterative methods have also been employed in the spherically symmetric setting, where it is observed that convergence is only achieved for a certain range of the parameters.
In fact there are indications that this convergent set corresponds with solutions which are \textit{dynamically} stable.
In particular, the highly relativistic multi-peak solutions mentioned above are not accessible with the iterative method.
An analysis of the fixed point map discussed above would provide useful insight on this observation -- see discussion in \Sectionref{sec.discussion.existence}.
For now, these numerical studies provide some evidence that the solutions obtained by the iteration procedure used in \cite{Ames2016,Ames2019} are dynamically stable.

\subsection{The axisymmetric Einstein--Vlasov System}
\label{sec.introduction.axisym_ev_system}
In order to review the topic and to describe open problems in a meaningful way we find it useful to formulate the system of equations and to introduce some of the quantities that we use to characterize the solutions. 
 
\subsubsection{Equations}
\label{sec.introduction.axisym_ev_system.equations}
The Einstein--Vlasov system consists of the coupled equations for the metric tensor $g$ and density function $f$ on phase space, which in arbitrary coordinates and geometric units ($G = c = 1$) reads
\[ Ric(g)_{ij} - \frac 12 R(g) g_{ij} = 8 \pi T(g,f)_{ij}, \quad p^i \partial_{x^i} f - \Gamma^{k}_{ij}(g) p^i p^j \partial_{p^k} f = 0, \]
where the second equation is called the Vlasov equation. 
Here $Ric(g), R(g)$ are the Ricci tensor and scalar of the metric $g$, $\Gamma^{k}_{ij}(g)$ are the Christoffel symbols of the metric $g$, and $T_{ij}(g,f)$ is the energy momentum tensor associated with the Vlasov matter. In a coordinate frame $(p^0,  p^1, p^2, p^3)$ on the tangent space at $x \in M$ the energy momentum tensor takes the form
\begin{equation}
T_{ij}(x) : = \int_{P_x} p_i p_j f(x,p)\frac{\sqrt{-\det g(x)}}{- p_0} \mathrm{d} p^1 \mathrm{d} p^2 \mathrm{d} p^3
\end{equation}
where $P_x$ is the mass shell at point $x$ and $p_0 = g_{0i} p^i$. 
For the stationary axisymmetric spacetimes considered in the papers \cite{Ames2016,Ames2019} the metric is  written in axial coordinates $(t, \rho, z, \phi)$ as
\begin{equation}
\label{eq:Metric}
g = - e^{2 \nu} dt^2 + e^{2 \mu} d\rho^2 + e^{2 \mu}dz^2 + \rho^2 B^2 e^{-2 \nu} (d\varphi- \omega dt)^2,
\end{equation}
where the metric fields $\nu, \mu, B, \omega$ depend only on the coordinates $\rho, z$. Note that $\rho = 0$ is the axis of symmetry, and that $(\rho, z)$ are cylindrical coordinates at infinity in the sense that in the appropriate limit $\rho$ is the radius of the symmetry group orbits. The metric field $\omega$ identically vanishes for solutions with no net rotation.

In order to construct stationary solutions it is useful to make an ansatz for the density function $f$ on phase space, namely we assume that $f$ depends only on the particle energy $E$ and angular momentum $L_z$ about the axis
\begin{equation}
\label{eq:AnsatzForm}
f(x,p) = K\Phi(E, L_z).
\end{equation}
Here $K$ is a positive constant, which we call the amplitude, and $\Phi$ is a given function. 
In principle the amplitude could be incorporated into $\Phi$ but since it plays an important role in the numerical fixed point methods used in \cite{Shapiro1993a,Shapiro1993b,Ames2016,Ames2019} we separate it out. 
The quantities $E$ and $L_z$ are given by
\begin{eqnarray*}
&& E =-g(\partial_t, p^i) = e^{2 \nu} p^0 + \omega (\rho B)^2 e^{-2\nu}  (p^3 - \omega p^0) \\
&& L_z = g(\partial_\phi, p^i) = (\rho B)^2 e^{-2 \nu} ( p^3 - \omega p^0),
\end{eqnarray*}
which are constant along the geodesic flow. As a consequence, when $f$ is given by the ansatz (\ref{eq:AnsatzForm}), the Vlasov equation is satisfied. 

In order to work with the integral expressions in the energy momentum tensor it is useful to introduce new momentum variables as follows
\begin{equation}
\label{eq:VBasis}
 v^0 = e^\nu p^0, \quad v^1 = e^\mu p^1, \quad v^2 = e^\mu p^2, \quad v^3 = \rho B e^{-\nu} ( p^3  - \omega p^0).
\end{equation}
In terms of these coordinates the energy momentum tensor can be written as
\begin{equation}
\label{eq:EMTensorVCoords}
T_{ij}(x) = \int_{P_x} p_i(v) p_j(v) f(x,v) \frac{\mathrm{d}^3 v}{\sqrt{1 + |v|^2}}.
\end{equation}
where the mass shell condition $g_{ij} p^i p^j = -1$, expressed in the new variables as $(v^0)^2 =  1 + |v|^2$, has been used to eliminate $p_0$. 
We take the positive root representing that all particles move forward in time. 
The particle angular momentum and energy can be expressed respectively as
\[  L_z =  \rho B e^{-\nu} v^3 =: \rho s \]
and
\[ E = e^\nu \sqrt{1 + |v|^2} + \omega L_z =: h + \omega \rho s. \]

By plugging in the ansatz for the density function in the expression for the energy momentum tensor the components become integral expressions in the metric fields. 
The EV system then takes the form a non-linear system of integro-partial differential equations which we state for completeness: 
\begin{align}
\label{eq:EinsteEqNU}
\Delta \nu & =
4 \pi \left(
\Phi_{00} + \Phi_{11}
+ \left( 1 + (\rho B)^2 e^{-4 \nu} \omega^2 \right)\Phi_{33}
+ 2 e^{-4 \nu} \omega \Phi_{03} \right)  \\\nonumber
& - \frac 1B \nabla B \cdot \nabla \nu
+ \frac 12 e^{-4\nu} (\rho B)^2 \nabla \omega \cdot \nabla \omega,  \\
\Delta B & = 8 \pi B \Phi_{11}  - \frac 1\rho \nabla \rho \cdot \nabla B, \\
\Delta \mu & =
- 4 \pi \left(
\Phi_{00} + \Phi_{11}
+ \left((\rho B)^2 e^{-4 \nu} \omega^2 - 1 \right)\Phi_{33}
+ 2 e^{-4 \nu} \omega \Phi_{03} \right)  \\ \nonumber
& + \frac 1B \nabla B \cdot \nabla \nu - \nabla \nu \cdot \nabla \nu
+ \frac 1\rho \nabla \rho \cdot \nabla \mu + \frac 1\rho \nabla \rho \cdot \nabla \nu \\
& + \frac 14 e^{-4\nu} (\rho B)^2 \nabla \omega \cdot \nabla \omega,   \\
\label{eq:EinsteEqWW}
\Delta \omega & = \frac{16 \pi}{(\rho B)^2} \left( \Phi_{03} + (\rho B)^2 \omega \Phi_{33} \right)
- \frac 3B  \nabla B \cdot \nabla \omega + 4 \nabla \nu \cdot \nabla \omega \\
& - \frac 2\rho \nabla \rho \cdot \nabla \omega,
\end{align}
where $\Delta u:= \rho^{-1} \partial_\rho(\rho \partial_\rho u) + \partial_z\partial_z u$ and $\nabla u = (\partial_\rho u, \partial_z u)$. The matter components are given by
\begin{align}
\label{eq:Phi00_expression}
\nonumber
\Phi_{00} 	&=  e^{2\mu - 2 \nu} T_{00} \\
		&= \frac{2\pi}{B} e^{2\mu - 2\nu} \int_{e^\nu}^\infty \int_{-s_l}^{s_l} E(h,s)^2 \Phi(E(h,s), \rho s) \mathrm{d}s \mathrm{d}h, \\
\nonumber
\Phi_{11} 	&= T_{\rho \rho} + T_{zz}  \\
		&= \frac{2\pi}{B^3} e^{2\mu + 2 \nu} \int_{e^\nu}^\infty \int_{-s_l}^{s_l} \left( s_l^2 - s^2 \right) \Phi(E(h,s), \rho s) \mathrm{d}s \mathrm{d}h, \\
\label{eq:Phi33_expression}
\nonumber
\Phi_{33} 	&= (\rho B)^{-2} e^{2\mu + 2 \nu} T_{\varphi \varphi} \\
		&=\frac{2\pi}{B^3} e^{2\mu + 2 \nu} \int_{e^\nu}^\infty \int_{-s_l}^{s_l} s^2 \Phi(E(h,s), \rho s) \mathrm{d}s \mathrm{d}h, \\
\label{eq:Phi03_expression}
\nonumber
\Phi_{03} 	&= e^{2\mu + 2 \nu} T_{0 \varphi} \\
		&= - 2\pi \rho B^{-1} e^{2\mu + 2 \nu}\int_{e^\nu}^\infty \int_{-s_l}^{s_l} s E(h,s) \Phi(E(h,s), \rho s) \mathrm{d}s \mathrm{d}h. 
\end{align}
Here 
\begin{equation}
\label{eq:SlDef}
s_l := B e^{-\nu} \sqrt{e^{-2\nu} h^2 -1} .
\end{equation}
There are also auxiliary equations which we choose not to write out but refer to \cite{Ames2016}. 

The system above must be complemented with boundary conditions. 
The following conditions guarantee that the solutions are asymptotically flat 
\[ 
\nu, \mu, \omega \to 0, \quad \text{and} \quad B \to 1, \quad \text{as} \quad r = |(\rho,z)| \to \infty. 
\]
In addition we require that the metric is regular at the axis which implies that
\begin{equation}
\label{eq:EinstBCAxis}
\nu(0, z) + \mu(0,z) = \ln B(0,z)
\end{equation}
for all $z$ in the solution domain. 

\subsubsection{Solution characteristics}
\label{sec.introduction.axisym_ev_system.solution_characteristics}

Our numerical solutions may be characterized by several quantities. 
In this review we only use a subset of these quantities and we refer to \cite{Ames2016,Ames2019} for a more complete picture. 

Two important properties of a solution are the total ADM mass $\mathcal{M}$ and the total angular momentum $\mathcal{J}$. 
We use the Komar expression for the mass and obtain 
\begin{equation}
\label{eq.MassAspect}
\mathcal{M} := 2\pi \int_{-\infty}^\infty \int_{0}^{\infty} g(\rho, z)\rho  \, \md \rho \dz,
\end{equation}
where
\[
g(\rho ,z)=e^{2\mu - 2\nu}\rho B T_{00} + \rho B (T_{\rho \rho} + T_{zz}) +  \frac{e^{2\mu + 2\nu}}{\rho B} T_{\varphi \varphi} - e^{2\mu - 2\nu} \rho B \omega^2 T_{\varphi \varphi}. 
\]
The mass plays an essential role in our iteration scheme. 
At each step of the iteration the ansatz function is renormalized such that the total mass is unity.
We also have a Komar integral expression for the total angular momentum
\begin{equation}
\label{eq:TotalAngularMomentum}
\mathcal J = -2\pi   \int_{-\infty}^\infty \int_{0}^{\infty} e^{2\mu - 2\nu} B  \left( T_{0\phi} + \omega T_{\phi \phi}   \right)\rho \, \md \rho \dz.
\end{equation}
The total angular momentum is particularly important when we construct highly compact solutions. 

An important measure of our solutions is the radius of support of the matter distribution. In spherical symmetry the ratio $2\mathcal M/R_0$, where $R_0$ is the radius of support in areal coordinates, is a measure of how relativistic a solution is. The ratio $2\mathcal M/R_0$ is often referred to as the compactness ratio. 

If we express the metric (\ref{eq:Metric}) in spherical coordinates, the radial coordinate $r$ is the isotropic radius. This can be related to the areal radial coordinate $R$ through
\[R = r ( 1+ \mathcal M/(2 r))^2.\]
In this paper we denote the areal radius of support by $R_0$ and the isotropic radius of support by $r_0$. 
For a spherically symmetric solution the radius of support can be determined from the cutoff energy $E_0$ (which is specified in the ansatz function) by the matching condition with a Schwarzschild exterior. The expression in terms of both the areal and isotropic coordinates is
\begin{equation}
\label{eq:E0andR0}
E_0 = \sqrt{1 - 2\mathcal M /R_0} = (1 - \mathcal M /(2 r_0))/ (1 + \mathcal M /(2 r_0)).
\end{equation}

In spherical symmetry a black hole forms if the mass $\mathcal M$ becomes confined within a Schwarzschild radius of $\mathcal R = 2 \mathcal M$. 
The compactness $2\mathcal M /R_0$ is thus a useful characterization of the solution. 
There is no such well-defined criteria in axisymmetry. In our setting, a natural measure of the radius is the length of the axisymmetric Killing vector field which we denote $\Rc := \rho B e^{-\nu}$. This quantity provides a natural length scale for the solution, in particular when restricted to the reflection plane ($z=0$) and evaluated near the boundary of the matter. For Vlasov matter, which typically has an extended atmosphere, it is useful to take the radius at which the compactness inside a cylinder of radius $\rho$ is maximum.
We define the compactness parameter $\Gamma := \max_{\rho \in (0, \infty)} 2 m(\rho)/\bar{R}_{\mathrm{circ}}(\rho)$, where $\bar{R}_{\mathrm{circ}} := (\Rc )|_{z=0}$ and where
\begin{equation}
\label{eq.MassAspect}
m(\rho) := 2\pi \int_{-\infty}^\infty \int_{0}^{\rho} g(\tilde{\rho}, z)\tilde\rho \, \md \tilde\rho \dz.
\end{equation}
Note that $m(\rho) = \mathcal M$ when $\rho$ exceeds the matter support. For the regular solutions we construct $\Gamma \in (0, 1)$.

\section{Brief survey of known results}
\label{sec.survey}
In this section we briefly summarize the current knowledge regarding stationary axisymmetric solutions of the Einstien-Vlasov system. 
Relevant details are discussed below in \Sectionref{sec.discussion} in the context of open problems.

\subsection{Numerical}
\label{sec.survey.numerical}

Stationary solutions to the axially symmetric Einstein-Vlasov system were for the first time constructed numerically in 1993 by Shapiro and Teukolsky in \cite{Shapiro1993a,Shapiro1993b}. 
In these works several sequences of stationary solutions are investigated with prolate and toroidal spatial density profiles, and include solutions with net total angular momentum. 
An exciting problem that these works left open was the question whether or not stationary solutions can be constructed which admit ergoregions.
In numerical work together with Anders Logg, the authors answered this question affirmatively in \cite{Ames2016}.
In this work stationary solutions with a variety of morphologies are generated, including disk-like, spindle-like, toroidal, and solutions formed from a composition of ansatz functions (multi-species solutions).
See \cite{Ames2023joss} for the numerical code used to generate these solutions.

Further investigations of highly rotating and relativistic sequences of toroidal solutions was carried out in \cite{Ames2019}. 
The numerical results suggest two distinct possible limiting spacetimes depending on whether the total angular momentum is larger or less than the mass squared. 
Sequences for which $\mathcal{J}$ becomes less than $\mathcal{M}^2$ eventually terminate, presumably becoming unstable to gravitational collapse. 
Sequences for which $\mathcal{J}$ stays larger than $\mathcal{M}^2$ approach what we dub the ``thin-ring limit'', for which the solutions appear to have properties similar to those of cosmic strings, in particular a locally conical geometry about the string characterized by a deficit angle. 
A bisection search is carried out that tunes between these two extremes, and the results suggest a possible quasistationary transition to an extremal Kerr black hole for the critical solution sequence.

In all of the numerical studies just cited the reduction scheme presented in \Sectionref{sec.introduction.axisym_ev_system.equations} and an ansatz of the form \Eqref{eq:AnsatzForm} is used.
The ansatz function $\Phi(E,L_z)$ is taken to have a product structure, i.e. 
\begin{equation}\label{eq.product_ansatz}
  \Phi(E,L_z)=\phi(E)\psi(L_z).
\end{equation}
In many cases the ansatz function for the energy is taken to be the polytropic one $\phi(E) = (E_0-E)_+^k$, where $(x)_+=x$ if $x\geq 0$ and $(x)_+=0$ if $x<0$, and $E_0, k$ are parameters.
The parameter $E_0$ has a natural interpretation as the cutoff energy for the particles, and its presence is important for obtaining solutions with compact support \cite{Rein2000}.
Different choices of $\psi$ result in different morphologies for the spatial density, among other properties.
We note that if $\psi$ is an even function the resulting solutions have zero net angular momentum.
If instead the ansatz function $\psi(L_z)$ vanishes for $L_z < 0$  then all particles have angular momentum of the same sign.
Solutions generated by such an ansatz have a net angular momentum and we call them rotating.

\subsection{Analytic}
\label{sec.survey.analytic}
In 2011 the existence of static axially symmetric solutions was shown for the first time by the second author together with Kunze and Rein in \cite{Andreasson:2011hg}.
In this case the total angular momentum vanishes.
This result was generalized in 2014 to the stationary, rotating case \cite{Andreasson2014}, and extended to the Einstein-Vlasov-Maxwell case (in which the particles have charge) in 2020 \cite{Thaller:2020}. 
These analytic results rely on a perturbation argument with the consequence that the constructed solutions can only be guaranteed to deviate slightly from spherically symmetric solutions and to have small total angular momentum.
Essentially the same system of equations as presented in \Sectionref{sec.introduction.axisym_ev_system} is used in the works \cite{Andreasson:2011hg,Andreasson2014}, with the technical difference that perturbation parameters corresponding to the angular momentum and relativistic nature of the solutions are introduced.
The authors also make use of an equation for $\xi := \nu + \mu$. 

In addition to the existence results discussed above, there is a recent result by Jabiri \cite{Jabiri2022} which uses a related method to construct stationary solutions.
The solutions are obtained as bifurcations from the Kerr spacetime and thus gives a generalization to the case where a black hole is surrounded by Vlasov matter.

We mention also that in the static and cylindrically symmetric setting (in which the solution has unbounded support in one-direction), the existence of solutions and compact support in the radial direction is proved in \cite{Fjallborg:2007}.

\section{Discussion and open problems}
\label{sec.discussion}
In this section we discuss properties of the solutions that were constructed analytically in \cite{Andreasson:2011hg,Andreasson2014} and numerically in \cite{Ames2016,Ames2019} with the aim of formulating open problems, both analytical and numerical. 
The first part concerns existence of analytic solutions and ideas on how to extend previous methods to more general ones.
The second part concerns highly relativistic compact solutions where ergoregions, the quasistationary transition to black holes and the thin ring limit are discussed.
The third part concerns models of galaxies where the aim is to find solutions with the morphology of galaxies as observed in nature.

\subsection{Existence of far-from spherically symmetric axially symmetric solutions}
\label{sec.discussion.existence}
A common feature of the existence results is that the solutions are obtained as perturbations of known solutions.
Hence solutions which are far-from spherically-symmetric in the sense of spatial density and net angular momentum --in particular, solutions containing ergoregions-- are not covered by these results.
Accordingly, there is extensive room for analytic progress on the existence of stationary solutions. 

One way to attack this problem is to mimic the numerical approach, i.e. to analytically investigate the iteration scheme of the numerical algorithm and show convergence in some domain of parameter space.
In fact there are several reasons why this may be of importance, some of which are discussed below.
To this end, let us describe the problem in the simplified case of the spherically symmetric Vlasov-Possion system.
This is the Newtonian analogue of the Einstein-Vlasov system. 
Although existence of static solutions to the Vlasov-Poisson system is well understood, it is nevertheless an interesting question if existence can be shown via the method suggested by the numerical algorithm. 

The Vlasov-Poisson system reads
\begin{eqnarray*}
  & &\partial_t f+v\cdot\partial_x f-\partial_x U\cdot\partial_v f=0,\\
  & & \Delta U=4\pi\rho, \;\; \lim_{|x|\to\infty}U(t,x)=0,\\
  & &\rho(t,x)=\int f(t,x,v)\, dv.
\end{eqnarray*}
This system has the same general structure as the Einstein-Vlasov system but it is much less involved. The aim is to prove the existence of static solutions by the following strategy. 
Let an ansatz $\Phi(E,L)$ for the particle distribution be given, i.e., 
\begin{equation} \label{ansatz}
f = \Phi(E,L).
\end{equation}
Here the particle energy $E$ and the modulus of angular momentum $L$ are given by
\[
E = \frac{1}{2} |v|^2 + U(x), \;\; L = |x\times v|^2.
\]
For a given spatial density $\rho = \rho(x)$ of finite mass $M>0$ and compact support we define its induced gravitational potential by 
\[
U_\rho(x) = -\int \frac{\rho(y)}{|x-y|} dy.
\]
If $U_\rho$ is substituted into the ansatz (\ref{ansatz}) we obtain
a new spatial density
\[
\tilde \rho (x) = \int \Phi\left(\frac{1}{2} |v|^2 + U_\rho (x), L\right)\, dv.
\]
We now define an amplitude
\[
K(\tilde \rho) = M \left(\int \tilde \rho(x)\, dx \right)^{-1}
\]
so that the new spatial density $K(\tilde \rho)\, \tilde \rho$ again
has mass $M$.
Let us consider the map $T:D\to D$ defined by $T(\rho)=K(\tilde{\rho})\tilde{\rho}$, where
\[
\tilde{\rho} = \int \Phi\left(\frac{1}{2} |v|^2 + U_\rho, L\right)\, dv.
\]
Here $D$ is a suitable domain. If this map has a fixed point $\rho^\ast$ then 
\[
\rho^\ast
= K(\rho^\ast)^{-1}
\int \Phi\left(\frac{1}{2} |v|^2 + U_{\rho^\ast}, L\right)\, dv,
\]
so that $(f^\ast,\rho^\ast,U_{\rho^\ast})$ is a static solution of
the Vlasov-Poisson system where $f^\ast$ is given by the new ansatz
\[
f^\ast = K(\rho^\ast)^{-1} \Phi(E,L).
\]
Hence the exact problem that is solved is a priori not known, it is determined once $ K(\rho^\ast)$ is known. It is an open problem to show that the map $T$ has a fixed point. 

Similar fixed point problems, based on the same strategy, can be formulated for several related systems; e.g. the spherically symmetric Einstein-Vlasov system and the axially symmetric Einstein-Vlasov system. It is an open problem in each case to show the existence of a fixed point. 

\subsubsection{Motivations}
\label{sec.disussion.existence.motivation}
One reason why this problem is important is obvious, it would give strong support that the numerical solutions obtained by this method are true solutions. Moreover, progress on the fixed point problem could give a new method to generate static solutions in cases where previous strategies have failed. We have not only in mind the axially symmetric Einstein-Vlasov system but also in the case of the Vlasov-Poisson system there is hope that this method can be used to construct new solutions, namely flat steady states. Presently there are only limited results in the literature about such solutions, cf. \cite{Rn1u2}, and they are of interest as models of disk galaxies, cf. \cite{Andreasson2015}. 

A further interesting aspect of this method is related to stability. Namely, as mentioned above, there are indications that solutions obtained by this algorithm are dynamically stable. This observation is particularly exciting in view of the open problem of non-linear stability for the spherically symmetric Einstein-Vlasov system. Hence, if a link between non-linear stability and the static solutions obtained by this algorithm can be established it would certainly be of great interest.

\subsection{Highly relativistic solutions}
\label{sec.discussion.properties}
In this section we discuss properties of highly relativistic solutions, by which we mean that the compactness ratio $\Gamma$ is large. 
Such solutions contain ergoregions resembling black hole solutions. 
A central open problem is whether or not a quasistationary transition to an extremal black hole is possible. This is discussed below. 

\subsubsection{Ergoregions}
\label{sec.disussion.properties.ergoregions}
An exciting result in \cite{Ames2016,Ames2019} was the discovery that there exist regular stationary toroidal solutions of the Einstein-Vlasov system which admit ergoregions (see for example Figures 3 and 4 in \cite{Ames2019}).
An ergoregion is typically associated with a Kerr black hole but not with a regular stationary solution.
The definition of an ergoregion is that the Killing field $\partial_t$ which corresponds to the time translation symmetry becomes spacelike.
In our parametrization of the metric it follows that an ergoregion is the set for which 
\begin{equation}\label{ergoregion}
  e^{2\nu}-\rho^2 B^2\omega^2 e^{-2\nu}<0. 
\end{equation}
In both \cite{Ames2016} and \cite{Ames2019} in which ergoregions were observed, the following ansatz function was used
\[
\Phi(E,L_z)=(E_0-E)_+^0(L_z - L_0)_+^0.
\]
This takes the form of \Eqref{eq.product_ansatz} with $k=0$, and $\psi(L_z)$ taking a similar polytropic form. 
Note however that the parameter $L_0$ represents a lower bound, resulting in a solution with net angular momentum.
In order to obtain sufficiently relativistic solutions we construct a sequence of solutions and ``gently" decrease the parameter $E_0$, seeding the solver for each new solution with the previous solution in the sequence. 

\begin{figure}[htb!]
  \centering
  \includegraphics[height=4cm]{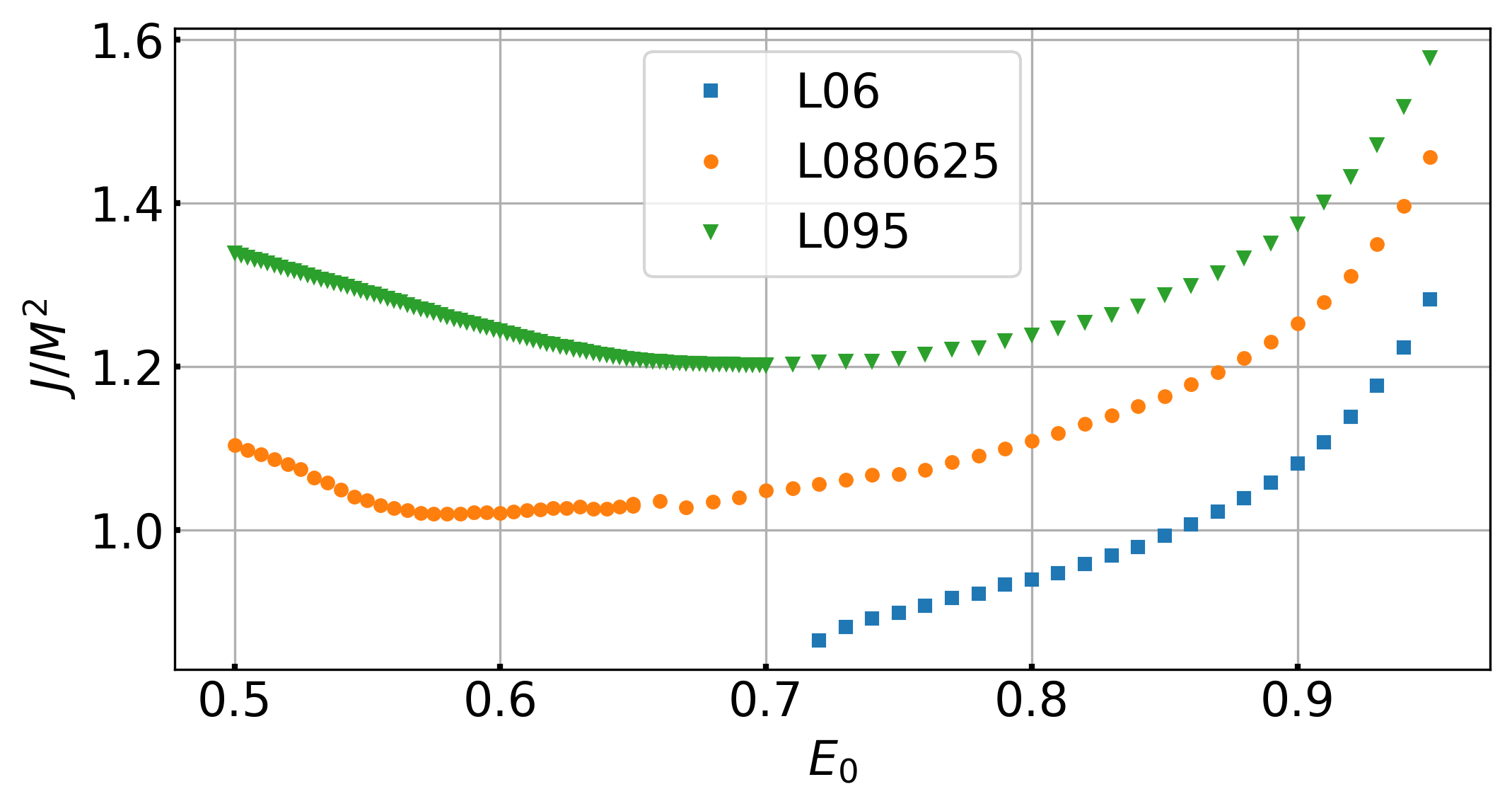}
  \captionof{figure}{
    Ratio of total angular momentum to total mass squared.
    Reproduced from Figure 2c in \cite{Ames2019}.
    }
  \label{fig.j_over_m_squared_vs_e0}
\end{figure}

The solutions admitting ergoregions that we obtain are all highly relativistic and highly rotating, each satisfying the inequality 
\begin{equation}\label{JM2}
  |\mathcal{J}|>\mathcal{M}^2.
\end{equation}
We note that the Kerr metric for which (\ref{JM2}) holds possesses a naked singularity. 
Hence, a stationary solution satisfying (\ref{JM2}) is likely stable (with respect to axially symmetric perturbations) in view of the weak cosmic censorship conjecture. 
A highly relativistic solution for which (\ref{JM2}) does not hold is most likely dynamically unstable; it will collapse to a Kerr black hole if it is perturbed, even in axisymmetry. 
This is consistent with the observation above in \Sectionref{sec.introduction.spherical_symmetry} that there are indications that our numerical algorithm only converges to dynamically stable solutions, and hence we are unable to obtain solutions with ergoregions for which $|\mathcal{J}|<\mathcal{M}^2$. 
The above discussion is illustrated in \Figref{fig.j_over_m_squared_vs_e0} where convergence is lost when $E_0<0.72$ for the sequence which does not satisfy the inequality (\ref{JM2}).
Perhaps a different numerical scheme will be more successful in finding solutions with ergoregions and with $|\mathcal{J}|<\mathcal{M}^2$.

As alluded to in \Sectionref{sec.survey.numerical}, Shapiro and Teukolsky also looked for solutions containing ergoregions in \cite{Shapiro1993b}.
They used a delta function based ansatz for both the particle energy and angular momentum,\footnote{Solutions obtained from distributions with imposed delta functions are typically not representative of general solutions to the Einstein-Vlasov system, see for example \cite{Ames2023}.} and obtained highly rotating and relativistic solutions.
At the limits of the resolution available at the time, Shapiro and Teukolsky were able to compute solutions corresponding to an $E_0$-value of about $0.745$.
In the step-function based ansatz studied in \cite{Ames2016,Ames2019}, solutions with an ergoregion appear in the sequences of solutions studied at around $E_0 \approx 0.66$.
Given this, one might speculate that the solutions obtained in \cite{Shapiro1993b} were not relativistic enough to contain ergoregions. 
It would be interesting to look for ergoregions in families of solutions obtained from other (non-step function) ansatzes.

\subsubsection{Quasistationary transition to black holes}
\label{sec.disussion.properties.quasistationary}
The presence of ergoregions suggests that the sequence of stationary regular solutions may be approaching a family of rotating  black hole solutions. Such a \emph{quasistationary transition} to black hole solutions does not occur for the spherically symmetric Einstein-Vlasov system due to a Buchdahl type inequality, which for a body of mass $\mathcal M$ and radius $\mathcal R$ reads $2 \mathcal M/ \mathcal R < 8/9$, cf. \cite{Andreasson:2008fu}. Hence there is a gap and $2 \mathcal M/ \mathcal R$ cannot be arbitrary close to one. However, if one allows for charge a similar bound relating the mass, radius, and total charge is known \cite{Andreasson:2008ge}, and in this case there is no gap; a quasistationary transition to a Reissner-Nordstr\"om black hole may thus be possible in spherical symmetry. Indeed, such a transition to the extremal Reissner-Nordstr\"om black hole has been shown by Meinel and H\"utten \cite{Meinel:cf} in the case of charged dust. Since charge is often considered as the poor man's angular momentum it is natural to ask if a similar transition is possible for rotating solutions. We note that in the case of disk solutions for dust, Meinel \cite{Meinel:2006eh,Meinel:2004hj,Meinel:2012tn} has answered this question affirmatively by analytic arguments. More general cases have been investigated numerically by Ansorg et al. \cite{Meinel:2012tn,Ansorg:2003dk,Fischer:2005bw}.

As discussed in \Sectionref{sec.survey.numerical} this question is investigated in \cite{Ames2019}. 
The angular momentum parameter $L_0$ is tuned between highly rotating solutions, and more slowly rotating solutions. 
For each value a sequence of solutions with gradually decreasing $E_0$ values is constructed, as shown in \Figref{fig.j_over_m_squared_vs_e0}.
The high $L_0$ solution sequences approach the thin-ring limit, while the low $L_0$ solution sequences eventually terminate, presumably becoming unstable to gravitational collapse.
Each such sequence, consisting of tens of solutions, is computationally expensive to complete. 
The most relativistic solutions we compute are with $L_0 = 0.80625$.
As shown in \Figref{fig.j_over_m_squared_vs_e0}, the ratio $\mathcal{J}/\mathcal{M}^2$ becomes very close to 1 as $E_0 \searrow 0.58$. 
Indeed, the compactness $\Gamma$ also obtains its maximum value of roughly $0.8$ at $E_0 = 0.58$, before decreasing again as the solution sequence bends towards the thin-ring limit (see \cite{Ames2019} Figure 2b). 
An extremal Kerr black hole has a compactness of $\Gamma = 1$. 

Given the presumed instability of solutions with $\J \sim \M^2$, and the conjectured relationship between stability and the iterative solution method, it is not surprising that obtaining solutions very near the extremal Kerr solution, even when $\J$ is just greater than $\M^2$, is challenging. With more computational effort and resolution can one continue this bisection search and reach all the way to a black hole solution? 
It is not clear that the answer to this question is affirmative. The numerical method used in the fluid case, cf. \cite{Meinel:2012tn,Ansorg:2003dk,Fischer:2005bw}, is different from the one used in \cite{Ames2019} and it is not straightforward to adapt that method to the Einstein-Vlasov system. 
Hence, from this discussion an essential question arises: is a transition of stationary solutions to black hole solutions possible?
Although such a transition has been established in the case of dust there is no \textit{a priori} requirement that a quasistationary transition to black hole solutions must occur for solutions of the Einstein-Vlasov system.

Since solutions of the Einstein-Vlasov system share many properties of solutions of field theoretical models such as the Einstein-Dirac system, cf. \cite{Andreasson2023:B}, we find it to be a question of fundamental importance to understand whether or not such a transition is possible. 
Perhaps there is a maximum value of the compactness parameter $\Gamma$ which is strictly below one as in the spherically symmetric case. 
Perhaps solutions necessarily approach the thin ring limit when $E_0$ is successively decreased independently of the value of $L_0$.
This would be surprising, since as mentioned above, static spherically symmetric charged solutions exist which are arbitrary close to black hole solutions, cf. \cite{Andreasson:2008ge,AndreassonEklundRein}. 
It may be that a different numerical algorithm is required to answer these questions or it may be that higher resolution is sufficient. 

\subsubsection{Existence of a thin-ring limit to sequences of rotating toroidal solutions}
\label{sec.discussion.properties.thin_ring}
The thin-ring limit discussed in the section above has interesting properties of its own. 
As shown in \cite{Ames2019} the limiting members of solution sequences approaching this limit appear to display a local conical geometry around the matter, reminiscent of cosmic strings. 
Due to the near Dirac nature of the spatial matter distribution in this limit, proving existence of such solutions may be analytically tractable. 
Indeed, in the spherically symmetric case it was utilized in \cite{Andreasson2007} that the highly compact solutions approached an infinitely thin shell where the matter sources become Dirac distributions.
This feature was essential for deriving upper and lower bounds on the compactness of the solutions, i.e. upper and lower bounds on $2m/r$. 
Perhaps a similar study is possible in the axially symmetric case by utilizing the Dirac nature of the thin ring solutions. 

\subsection{Models of galaxies}
One goal in \cite{Ames2016} was to construct solutions which resemble 
galaxies as observed in nature. 
While the solutions obtained in \cite{Ames2016} make no attempt to be full-fledged galaxy models, it is still of interest to compare their properties with real galaxies in order to better understand which properties of galaxies are captured by the fundamental axially symmetric gravitational physics. 

We point out that models of galaxies are most often obtained within the Newtonian framework, cf. \cite{BT} for models based on the Vlasov-Poisson system. One reason why relativistic models could be of interest in this context is that net rotation has no effect on the gravitational field in Newtonian gravity, yet it noticeably alters the geometry in general relativity, even for solutions which are not very relativistic.

\subsubsection{Solutions with disk-like morphology}
\label{sec.discussion.properties.disk_like}
An important class of galaxies are disk galaxies such as the Milky Way.
Given the prevalence of disk galaxies in the observable universe, it is reasonable to ask whether such a morphology is represented in the space of stationary solutions of the Einstein-Vlasov system. 
There exist models of disk galaxies which are confined to the plane, cf. e.g. \cite{Schenk1998,Andreasson2015}, but it would be desirable to obtain three-dimensional solutions which are disk-like in the sense that the core of the density is as close to planar as possible. 

An ansatz in which the z-component of the angular momentum is taken to have a Gaussian profile generates oblate spheroidal solutions. 
The ansatz is given by
\begin{equation}
\label{eq:GaussianLAnsatz}
  \psi(E,L_z) = \frac 1L_0 \exp(L_z^2/L_0^2).
\end{equation}
This ansatz is symmetric in $L_z$ and therefore generates non-rotating solutions of the Einstein-Vlasov system. 
A rotating version can be constructed by additionally imposing that all particles have $L_z$ of the same sign.
In the limit $L_0 \to \infty$ the ansatz becomes independent of $L_z$, thus generating a spherically symmetric spatial density. 
As $L_0$ is decreased, particles with higher angular momentum are more heavily weighted compared to those with low angular momentum. 
This yields a more flattened shape. 

\begin{figure}[htb!]
  \centering
  \includegraphics[width=.75\linewidth]{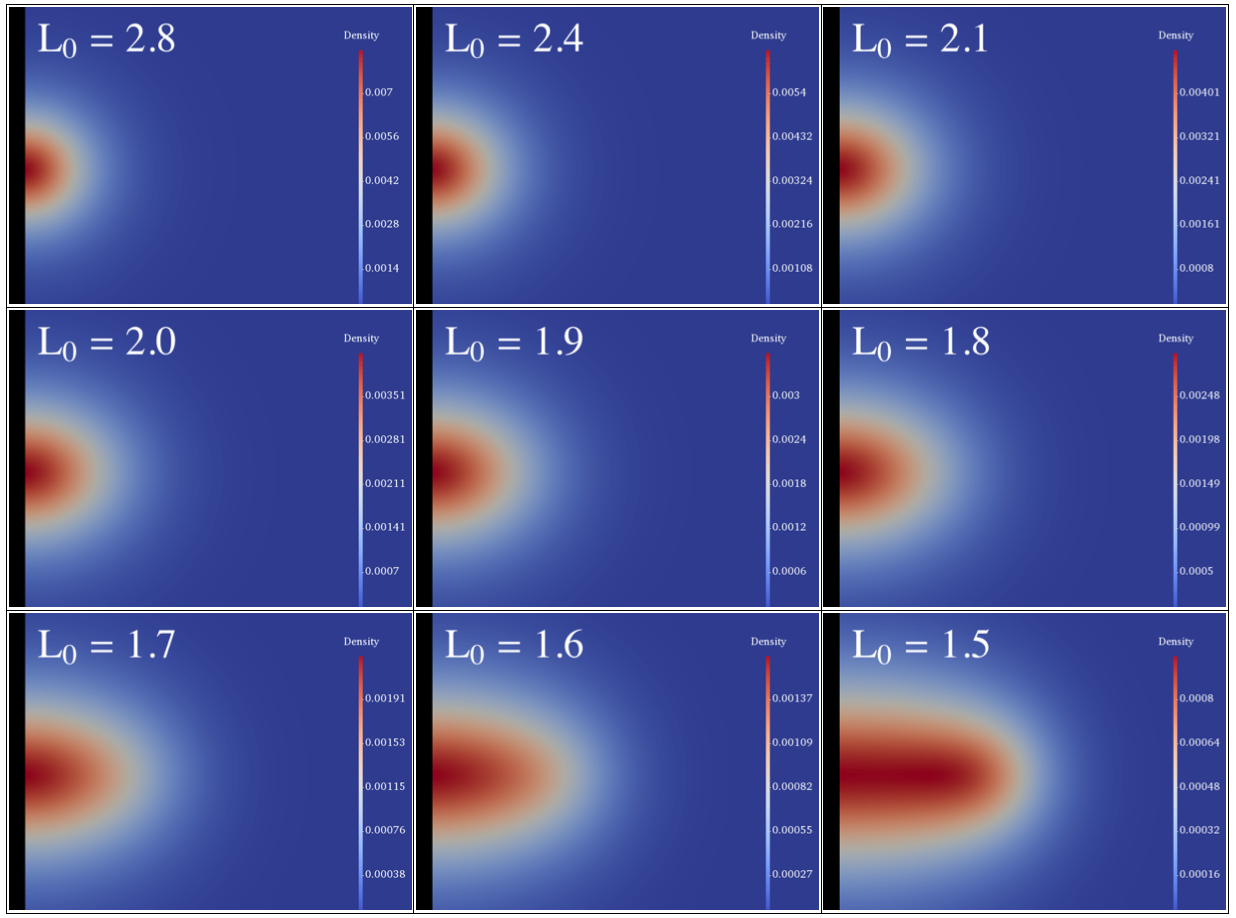}
  \captionof{figure}{
    $L_0$-parameterized sequence of disc-like solutions in the Einstein-Vlasov model.
    Reproduced from \cite{Ames2016} Figure 8.}
  \label{fig.ev_disc_family}
\end{figure}

In \cite{Ames2016} the flattening of such non-rotating oblate spheroids is studied.
Parameters $k$ and $E_0$ for the particle energy are fixed, while the parameter $L_0$ is gradually reduced from $10$ to $1.5$. 
\Figref{fig.ev_disc_family} shows the spatial density for a selection of solutions.
Within the parameter range shown the spatial density distribution stretches to its most flattened form, while for parameters $L_0 < 1.5$ our numerical algorithm does not converge.
We interpret this lack of convergence as a failure of the configuration to remain gravitationally bound. 
The final solution in the sequence does not have a very flattened form, though higher density contours appear far more disk-like than lower density contours -- see also Figures 9 and 10 in \cite{Ames2016}. 

In \cite{Ames2016} models with net angular momentum are also studied. It is clear that rotation has an effect on the geometry, and hence the morphology, but it does not give rise to the very flattened form that we look for (see for example Figure 9 of \cite{Ames2016}). It should be stressed that a systematic investigation of this effect was not carried out in \cite{Ames2016} but it would be surprising if merely an increase of the angular momentum would result in sufficiently flattened solutions. 

In conclusion, our solutions do not satisfyingly resemble the extremely flattened galaxies which are observed in nature. We find it to be a central open problem to answer the question whether or not it is possible to obtain solutions of the Einstein-Vlasov system which resemble realistic disk galaxies. Perhaps a different choice of ansatz function will work out, or a different numerical method, or that it is simply not possible. It would be a great contribution to find out the answer to this question. 

A different aspect is that it is often claimed that halos of dark matter may be crucial to the pronounced flattened shape of disk galaxies. It is a question of great importance to investigate if such halos surrounding disk-like solutions do improve the convergence of the numerical algorithm. If this turns out to be the case, it would give significant support to the established claim that disk galaxies are embedded in dark matter halos. To our knowledge this has not been investigated for the Einstein-Vlasov system and we find it to be an essential open problem.

\subsubsection{Spindle-like, toriodal and composite solutions}
\label{sec.discussion.properties.galaxy_morphology}
In \cite{Ames2016} we also construct spindle-like and toroidal solutions similar to what was done in \cite{Shapiro1993b}. 
The ansatz function we use for spindle solutions is given by
\begin{equation}
\label{eq:AndreassonLAnsatz}
\psi_{spindle}(L_z) = \begin{cases}
	(1 - Q |L_z| )^l,  |L_z| < 1/Q \\
       0,  		         |L_z| \geq 1/Q ,
   \end{cases}
\end{equation}
and the ansatz function for the toroidal solutions takes the form 
\begin{equation}
\label{eq:toroidal}
\psi_{torus}(L_z) =
\begin{cases}
       (|L_z| - L_0)^l,  |L_z| > L_0 \\
       0,  		     	|L_z| \leq L_0. \\
   \end{cases}
\end{equation}
Here $Q,L_0$ and $l$ are parameters. 
As a remark we mention that the ansatz (\ref{eq:AndreassonLAnsatz}) was used for the Vlasov-Poission system in \cite{Andreasson2015} to investigate the rotation velocities of stars in galaxies. 
In that setting, this ansatz gives rise to flat rotation curves similar to those found in observations.

Interestingly, astrophysical objects with spindle-like and toroidal structures have recently recieved attention by astrophysicists. 
In 2017 galaxy surveys revealed that prolate spindle-like galaxies are much more common than previously thought \cite{Tsatsi2017}. 
In 2019 it was announced that the VLA telescope had directly imaged a toroidal structure within an active galactic nucleus \cite{Carilli2019}. 
In view of the observational evidence of these types of objects we find that a more careful study of spindle solutions and toroidal solutions is motivated, where the features of the numerical solutions should be compared with the features of the observed objects. 

In the context of galaxy morphology, let us also discuss composite objects which are obtained by combining different ansatz functions. 
Examples of composite astrophysical objects are numerous, and include disk galaxies with a central bulge, galaxies with dark matter halos, as well as ring galaxies. 
In the case of the Vlasov-Poisson system there are several results in the literature about composite solutions, but for the Einstein-Vlasov system they were first obtained in \cite{Ames2016}. 

\begin{figure}[htb!]
\centering
  \includegraphics[width=.5\linewidth]{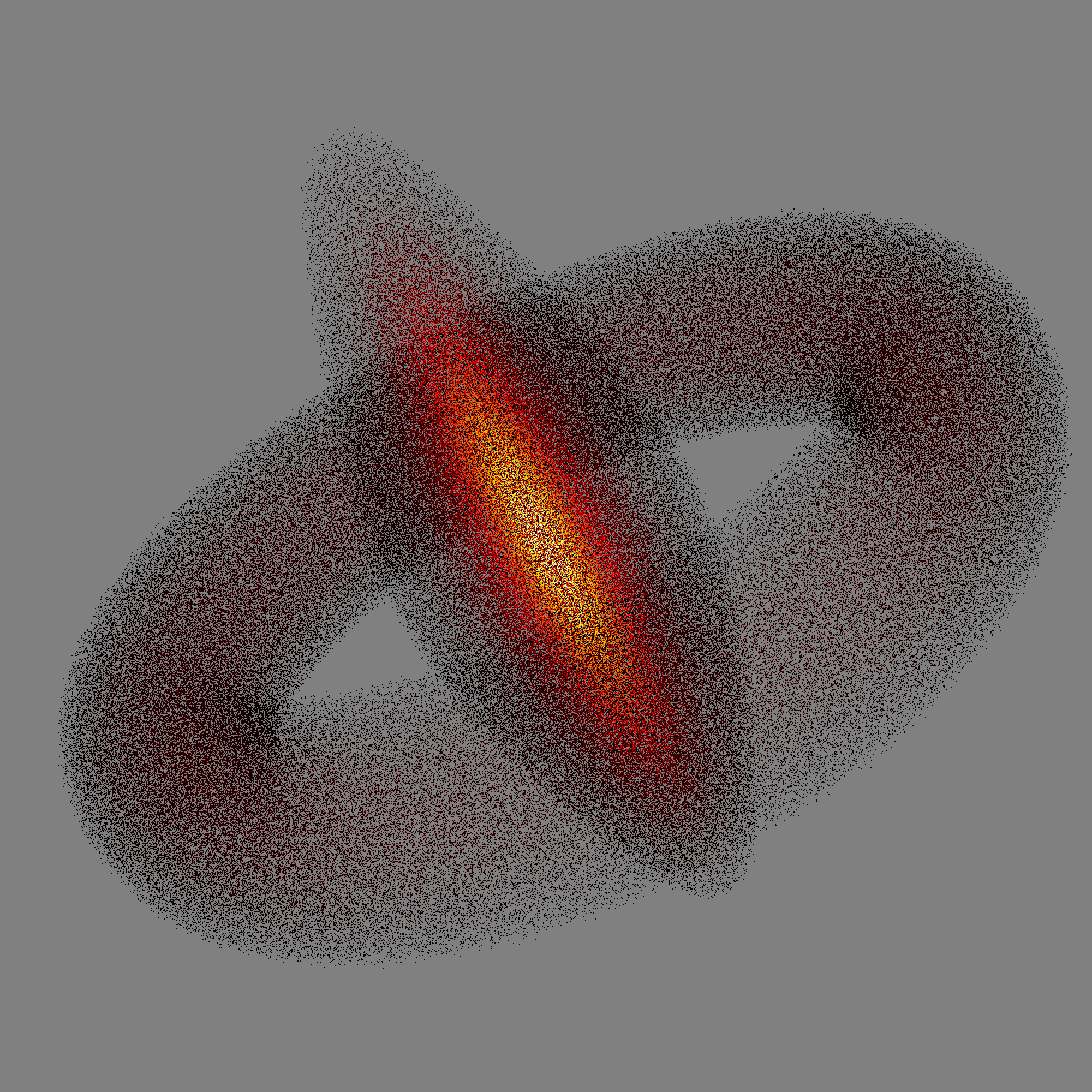}
  \captionof{figure}{Point cloud representation of a composite solution formed from spindle (\ref{eq:AndreassonLAnsatz}) an torus (\ref{eq:toroidal}) ansatzes, and inspired by Hoag's object \cite{Hoag1950}}
  \label{fig:spindlehoag_fancy}
\end{figure}

It turns out to be sensitive to combine ansatz functions. 
Our numerical algorithm does not converge for an arbitrary non-trivial linear combination of ansatz functions even if they individually give rise to solutions. 
For instance, we are not able to obtain convergence by combining an ansatz for a polytropic central bulge with a toroidal ansatz. 
The spindle solutions turned out to be useful in constructing composite objects. 
In \cite{Ames2016} we used the following form of ansatz function for a composite solution 
\[\Phi(E, L_z) = K_s \Phi_{spindle}(E,L_z) + K_t \Phi_{torus}(E,L_z),\]
where $K_s$ and $K_t$ are amplitudes, or weights, of the two ansatz functions. 
An example of a solution inspired by Hoag's object \cite{Hoag1950} and obtained in this way is shown in Figure \ref{fig:spindlehoag_fancy}. 
An open issue is to understand which combinations of ansatz functions give rise to composite solutions and to find out if these solutions resemble the morphology of composite objects found in nature.

\subsection{The massless Einstein-Vlasov system}
\label{sec.discussion.massless}
A question related to the opening of this review and to the existence of compact solutions is whether or not massless axially symmetric solutions exist.
In the spherically symmetric case massless solutions exist if the solutions are sufficiently compact, cf. \cite{Andreasson:2016jo,Andreasson2021}. Indeed, highly compact spherically symmetric massless solutions have the property that the density function vanishes at some finite radius $R$ and the solution can be glued to a vacuum Schwarzschild solution at $r=R$. 
Without the gluing procedure the density function $f$ will eventually become strictly positive at a sufficiently large radius, cf. \cite{Andreasson:2016jo}. This particular feature is only present in the massless case. Hence, the crucial question is whether or not massless axially symmetric solutions can be constructed such that the density function vanishes outside a compact set. 
If so, it would be possible to glue the solution to a vacuum solution --although no explicit vacuum solution exists as in the spherically symmetric case. 

We find this question exciting in view of the quote by Wheeler given in the introduction. 
Wheeler wanted to find and investigate the properties of regular solutions of the Einstein-Maxwell system. 
He named such solutions \textit{geons}. 
In spherical symmetry the only solution to the Einstein-Maxwell system is the Reissner-Nordstr{\"o}m solution which is not regular. 
Hence, spherically symmetric geons do not exist strictly speaking, although Wheeler did introduce the concept of \textit{idealized} spherically symmetric geons \cite{Wheeler1955}. 
His ultimate wish was to study the axially symmetric case. 
Since an electromagnetic field can be modeled as a photon gas, a possible way to obtain understanding of the properties of solutions of the Einstein-Maxwell system is to study solutions of the massless Einstein-Vlasov system. 
In the spherically symmetric case highly relativistic solutions of the two systems are indeed very similar, cf. \cite{Andreasson:2016jo}. 
Hence, the question whether or not axially symmetric solutions of the massless Einstein-Vlasov system exist is closely related to the existence of Wheeler's original conception of geons.

\subsection{Regular solutions about central black holes}
\label{sec.discussion.solutions_with_bh}
An interesting direction in which to extend the study of regular axisymmetric stationary solutions of the Einstein-Vlasov system, is to consider solutions with a central black hole. 
This imposes inner boundary conditions at the black hole horizon. 
One motivation is to better understand coupled matter-black hole systems and the effects that matter can have on central black holes.
Another reason such a line of investigation is of interest is to provide foundational results for current research on accretion disks. 
To date and to the author's knowledge, such studies make use of a fixed black hole background solution, as in for example \cite{Parfrey2019}, and focus on the plasma physics in this extreme gravitational environment. 
In the purely gravitational setting, Rioseco and Sarbach have several works studying the dynamics of Vlasov matter on black hole backgrounds, for example \cite{Rioseco2018,Rioseco2023}.

The self-gravitating case is much more challenging.
In the spherically symmetric setting, Andr{\'e}asson proves the existence of solutions with a central Schwarzschild black hole in the massless case \cite{Andreasson2021}, and in \cite{Rein1994} Rein establishes a such existence in the massive case.
As mentioned in \Sectionref{sec.survey.analytic}, Jabiri has proved existence of self-gravitating solutions using a perturbation argument about the Kerr spacetimes \cite{Jabiri2022}. 
This results represents the first proof of existence for axially symmetric self-gravitating solutions about a central black hole. 
By the nature of the proof however, the matter must be small (close to vacuum), and it remains an open problem to find general axially symmetric self-gravitating solutions of the Einstein-Vlasov system with a central black hole. 

A sensible starting place is to probe this problem numerically. 
One approach would be to modify the code used in \cite{Ames2016,Ames2019} and documented in \cite{Ames2023joss}.
Related studies exist in the case of uniformly rotating fluids about black holes, see for example \cite{Bardeen1973,Ansorg2005}.

\subsection{Extension to the Einstein-Vlasov-Maxwell system}
\label{sec.discussion.einstein_vlasov_maxwell}
Another interesting direction of research, and possible extension of the code \cite{Ames2023joss}, is the addition of charge. 
Allowing the particles to be charged (in addition to having mass) and interact additionally through the electromagnetic field leads to the Einstein-Vlasov-Maxwell system. 
This system was studied numerically in 2009 by Andr{\'e}asson and coauthors \cite{AndreassonEklundRein} in the spherically symmetric setting, and in 2020 Thaller used perturbation methods to prove existence of solutions in the axisymmetric setting \cite{Thaller:2020}. 
This result is similar to the work of \cite{Andreasson2014} except that the reference solution is a charged solution of the spherically symmetric Vlasov-Poisson system, and as a result the particle charge is not restricted to be small. 
The total angular momentum and the strength of the gravitational field are still restricted, and an understanding of general solutions to the axially symmetric Einstein-Vlasov-Maxwell system is lacking.

\section{Remarks on the evolution problem}
\label{sec.evolution}
We have reviewed the status of \textit{stationary} solutions of the axially symmetric Einstein-Vlasov system.
The question of their stability, and more generally, the fate of any axially symmetric initial data requires an evolution code.
Shapiro and Teukolsky were pioneers in developing such a code using the particle in cell (PIC) method.
More recently, several results on the evolution problem have been carried out, cf. \cite{Yoo2017,East2019,Ames2021,Ames2023}. 
We have in this work left out a discussion about these results.
One reason being that, to our knowledge, no evolution code is open source and it is an extensive work to develop such a code.
Needless to say there is immense room for improvements and developments of this topic.
We will not enter into this here but let us at least bring up one open problem which is related to the discussion above.
The highly compact solutions that were constructed in \cite{Ames2019} require very high resolution.
It is an outstanding problem to determine whether or not these solutions are stable.
The high resolution needed to construct the stationary solutions is clearly a severe obstacle.
How can the high resolution needed for the stationary solution be carried over to the evolution problem to make a simulation practically feasible?

\section{Acknowledgements}
The second author acknowledges support from the Erwin Schr{\"o}dinger Institute where parts of this work was carried out during the program ``Spectral Theory and Mathematical Relativity", in June 2023.

%%%%%%%%%%%%%%%%
%%%%%%%%%%%%%%%%
\bibliography{references}
\end{document}